\documentstyle[12pt,epsf,psfig]{article}
\begin{document}

\begin{center}
{\Large\bf Dynamic structure factor of gold } \\

\vspace{0.5cm}
{  I. G. Gurtubay,$^{\rm (1)}$
J. M. Pitarke,$^{(1,2)}$
 I. Campillo$^{(1)}$
and 
A. Rubio$^{(3)}$}\\

\vspace{0.6cm}
{\it $^{\rm (1)}$ Materia Kondentsatuaren Fisika Saila, Zientzi Fakultatea,
Euskal Herriko Unibertsitatea, 644 Posta kutxatila, 48080 Bilbo, Basque Country,
Spain }\\
{\it $^{\rm (2)}$ Donostia International Physics Center (DIPC) and Centro Mixto
CSIC-UPV/EHU, Donostia, Basque Country, Spain }\\
{\it $^{\rm (3)}$ Departamento de F{\'{\i}}sica Te\'orica, Universidad de 
Valladolid, 47011 Valladolid, Spain }\\
\end{center}

\begin{abstract}
We have investigated the role of localized {\it d} bands in the dynamical
response of Au, on the basis of {\it ab initio} pseudopotential calculations. 
The density-response function has been evaluated in the random-phase
approximation. For small momenta, we have found 
a double peak structure in the energy-loss
function, which results from the presence of  
{\it d} electrons.
 These results are in good agreement with the experimentally determined optical
response of gold. We also analyze the dependence of the dynamical structure
factor on the wave vector {\bf q}.

\end{abstract}

\vspace{0.2cm}

\section{Introduction}
Many experimental and theoretical work
\cite{Ehr}-\cite{Cazalilla}
 have focused on noble metal systems
 in order to get a better understanding of the role played by localized
 $d$ electrons on the properties of delocalized, free-electron-like, electrons.

The behavior of noble metals has been considered by Ehrenreich and Philipp
\cite{Ehr}.
These authors used the data obtained in experimental measurements of the
 reflectivity from Cu and Ag \cite{Taft} to obtain the optical  
spectra. They separated  the Drude and interband contributions to
the optical response, explaining why the energy of the plasma resonance
in Ag is shifted down in energy.
In a later work, Cooper, Ehrenreich and Philipp \cite{Cooper} extended 
this analysis to Au.

Gold and copper present no decoupling between 
$sp$ and $d$ orbitals, 
and a combined description of these one-electron states is needed to address
 both structural and electronic properties of these  materials.

Due to the rapid progress in computing technology, realistic calculations of 
the density-response function of solids 
have become feasible in the past few years.
Recently, the dielectric matrix as well as the energy-loss spectrum have been
evaluated numerically within a first principles framework for several simple
(Al \cite{AL}, Cs \cite{CS}, Be \cite{BE} ) and noble 
\cite{Igor}-\cite{Cazalilla} metals, 
as well as for some insulators
like Si \cite{SI}, LiF \cite{LIF} or Al$_2$O$_3$ \cite{CHIN}. 
In the case of noble metals, the energy-loss function for Cu has been 
evaluated in Ref. 4 
and plasmonic excitations
in Ag have been reported in Ref. 5.
Within the same many-body framework, {\em ab initio} calculations of the 
electronic stopping power of real solids \cite{stop} and excited electron 
lifetimes \cite{Igor2} have also been reported.

In this paper we report an {\em ab initio} evaluation of the dynamical 
response function of Au as computed in the random-phase approximation 
(RPA) after an 
expansion of all 6$s^1$ and 5$d^{10}$ one-electron states in a plane wave basis.

\section{Formalism}

The linear density-response function $\chi({\bf r},{\bf  r'};\omega)\ $ of an
interacting electronic system is defined by the equation 
\begin{eqnarray}\label{eq1}
\rho^{ind}({\bf r},\omega)=\int{\rm d}{\bf r}'\,
\chi({\bf r},{\bf  r'};\omega)\,V^{ext}({\bf r'},\omega),
\end{eqnarray}
where $\rho^{ind}({\bf r},\omega)$ is the electron density induced by an 
external potential $V^{ext}({\bf r'},\omega)$ \cite{Pines}.

For periodic crystals we Fourier transform $ \chi({\bf r},{\bf r}';\omega) $ 
into a matrix
$\chi_{{\bf G},{\bf G}'}({\bf q},\omega)$ which, in the framework of 
time-dependent density-functional theory (TDDFT) \cite{Dreizler}, 
satisfies the matrix equation
\begin{eqnarray}\label{eq2}
\chi_{{\bf G},{\bf G}'}({\bf q},\omega)=\chi^0_{{\bf G},{\bf G}'}({\bf
q},\omega)+\sum_{{\bf G}''}\sum_{{\bf G}'''}\chi^0_{{\bf G},{\bf G}''}({\bf
q},\omega)\cr\cr
\times\left[v_{{\bf G}''}({\bf q})\delta_{{\bf G}'',{\bf G}'''}+K^{xc}_{{\bf
G}'',{\bf G}'''}({\bf q},\omega)\right]
\chi_{{\bf G}''',{\bf G}'}({\bf q},\omega).
\end{eqnarray}
\\
The wave vector ${\bf q}$  is in the first Brillouin zone (BZ), ${\bf G}$
 and ${\bf G}'$ are reciprocal lattice vectors, $v_{\bf G}({\bf q})=4\pi/|{\bf
q}+{\bf G}|^2$ are the Fourier coefficients of the bare Coulomb potential, the
kernel $K^{xc}_{{\bf G},{\bf G}'}({\bf q},\omega)$ accounts for short-range
exchange-correlation effects (in the RPA, 
$K^{xc}_{{\bf G},{\bf G}'}({\bf q},\omega)$ = 0)
and $\chi^0_{{\bf G},{\bf
G}'}({\bf q},\omega)$ are the Fourier coefficients of the density-response
function of noninteracting Kohn-Sham electrons:
\begin{eqnarray}\label{eq3}
\chi_{{\bf G},{\bf G}'}^0({\bf q},\omega)={1\over \Omega}\sum_{\bf
k}^{BZ}\sum_{n,n'} {f_{{\bf k},n}-f_{{\bf k}+{\bf q},n'}\over E_{{\bf
k},n}-E_{{\bf k}+{\bf q},n'} +\hbar(\omega + {\rm i}\eta)}\cr\cr
\times\langle\phi_{{\bf k},n}|e^{-{\rm i}({\bf q}+{\bf G})\cdot{\bf
r}}|\phi_{{\bf k}+{\bf q},n'}\rangle
\langle\phi_{{\bf k}+{\bf q},n'}|e^{{\rm i}({\bf q}+{\bf G}')\cdot{\bf
r}}|\phi_{{\bf k},n}\rangle,
\end{eqnarray}
\\
where the second sum runs over the band structure for each wave vector
${\bf k}$ in the first BZ, $f_{{\bf k},n}$ represents the Fermi-Dirac 
distribution function,  $\eta$ is a positive infinitesimal,  and $\Omega$ 
represents the normalization volume.  $\phi_{{\bf k},n}({\bf r},\omega)$ and
$E_{{\bf k},n}$ are the Bloch eigenfunctions and eigenvalues of the
Kohn-Sham Hamiltonian of density-functional theory (DFT) \cite{Kohn}. 

The inverse dielectric function is connected to the density-response 
function   by the following relation:
\begin{equation}
\epsilon_{{\bf G},{\bf G}'}^{-1}({\bf q},\omega)=\delta_{{\bf G},{\bf G}'}+v_{{
\bf
G}'}({\bf q})\chi_{{\bf G},{\bf G}'}({\bf q},\omega).
\end{equation}
\\
The properties of the long-wavelength limit (${\bf q}\to 0$) of the dynamical
density-response function are accessible by measurements of the optical
absorption, through the imaginary part of the dielectric response function
$\epsilon_{{\bf G}=0,{\bf G}'=0}({\bf q}=0,\omega)$.

For the evaluation of the one-electron Bloch states, we first expand them 
in a plane-wave basis,
\begin{equation}\label{eq10}
\phi_{{\bf k},n}({\bf r})={1\over\sqrt V}\sum_{\bf G} u_{{\bf k},n}({\bf
G}){\rm e}^{{\rm i}({\bf k}+{\bf G})\cdot{\bf r}},
\end{equation}
\\
with kinetic-energy cutoff of 75 Ry ($\sim 10^4\,{\bf G}$-vectors) and then 
evaluate the Kohn-Sham equation of DFT with a full description of the 
electron-ion interaction based on the use of a non-local, 
norm-conserving ionic pseudopotential \cite{Hamman}. Finally, we evaluate from 
Eq. (\ref{eq3}) the Fourier coefficients 
$ \chi_{{\bf G}, {\bf G}'}^0({\bf q},\omega)$, and solve the matrix equation,
Eq. (\ref{eq2}),
for the Fourier coefficients of the interacting density-response function, 
which we evaluate
in the RPA. 

\begin{figure}[t]
\epsfclipon
\epsfxsize=8cm 
\centerline{\psfig{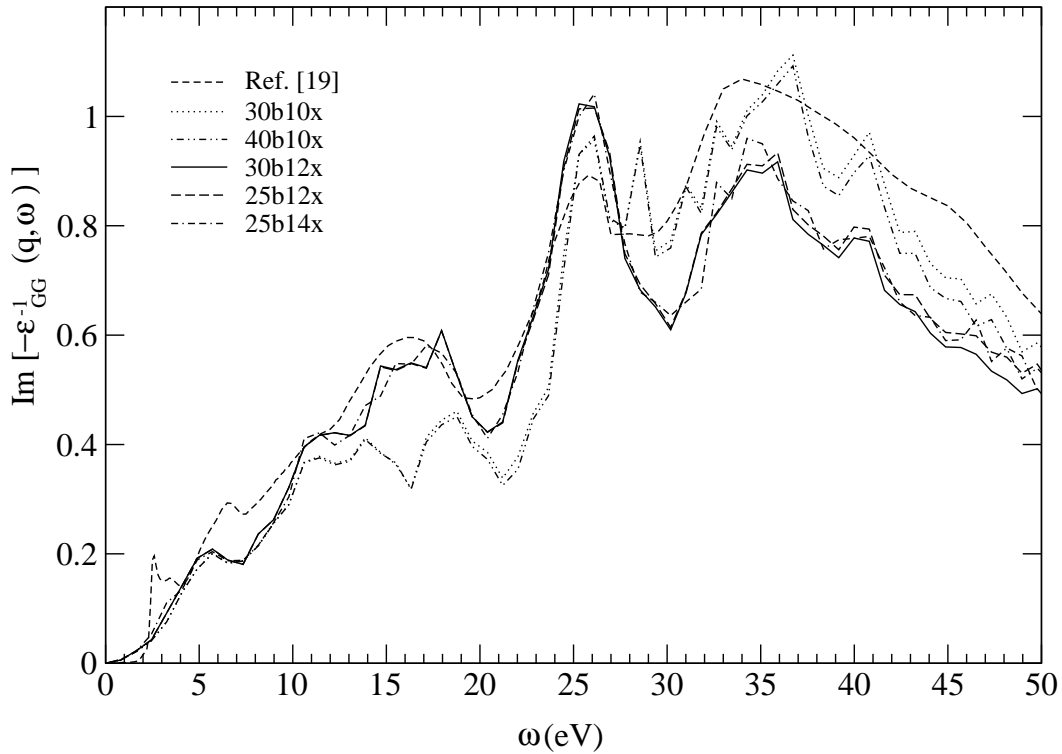}}
\caption{\protect\small 
Convergence for the energy-loss function for different number of bands and
samplings for 
${\bf q}=0.16(1,0,0)\times(2\pi)/a$ ($a=3.01\,\AA$ is the experimental
lattice constant) and ${\bf G}=0$. Optical data are from Ref. 19. 
"b" stands for bands and "$\times$" for the sampling (See the text 
for reference).}
\label{fig:conv}
\end{figure}

We have performed calculations for the energy-loss function using 
25, 30 and 40 bands, and samplings over the BZ on $10\times 10\times 10$, 
$12\times 12\times 12$ and $14\times 14\times 14$ Monkhorst-Pack meshes
\cite{MP}. 25 bands and a $12\times 12\times 12$ sampling
turned out to be enough to reach convergence (See Fig. \ref{fig:conv}), 
but 30 bands were used to 
guarantee good values for the highest energies under study.

\section{Results}
\label{Results}

\begin{figure}[t]
\epsfclipon
\epsfxsize=8cm 
\centerline{\psfig{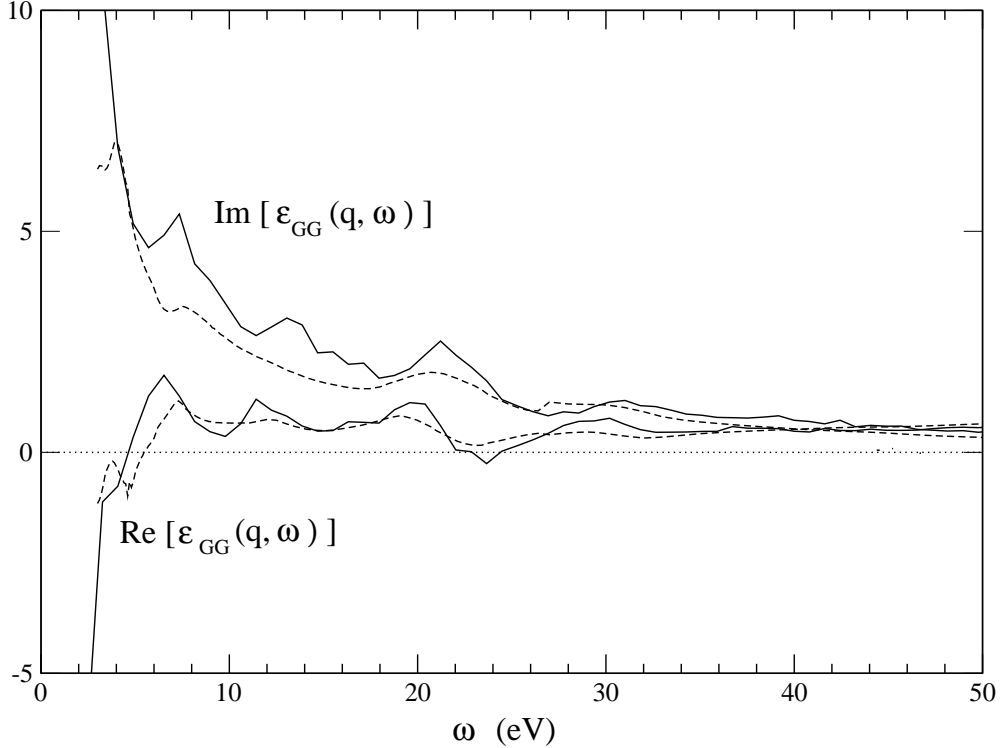}}
\caption{\protect\small 
Real and imaginary parts of the  $\epsilon_{{\bf G},{\bf G}}({\bf
q},\omega)$ dielectric function of Au, for ${\bf
q}=0.16(1,0,0)\times(2\pi)/a$ ($a=3.01\,\AA$ is the experimental
lattice constant) and ${\bf G}=0$. Solid and dashed lines represent
our calculations and the optical data of
Ref. 19, respectively.}

\label{fig:epsilon}
\end{figure}

Fig.\ \ref{fig:epsilon} shows our \em{ab initio} \rm calculations (solid line) 
for both real and imaginary parts of the 
$\epsilon_{{\bf G}=0,{\bf G}'=0}({\bf q}=0,\omega)$
dielectric function of gold for a small momentum transfer of 
$|{\bf q}+{\bf G}|=0.13\,a_0^{-1}$ ($a_0$ is the Bohr radius) 
together with the optical data
(${\bf q}=0$) of Ref. 19 (dashed line). 
%
%

\begin{figure}[t]
\epsfclipon
\epsfxsize=8cm 
\centerline{\psfig{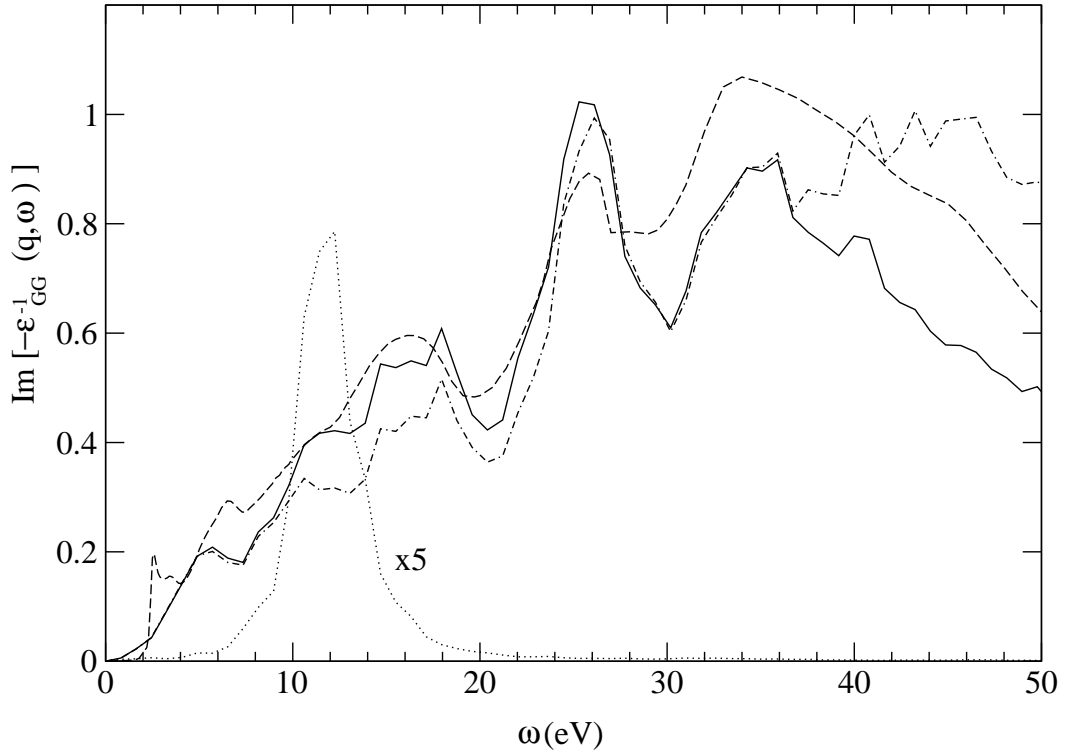}}
\caption{\protect\small  
As in Fig.\ \ref{fig:epsilon} for the energy-loss function of Au.
The dotted line represents the result of assigning the
$5d$ shell to the core and the dashed-dotted line the result of
excluding the crystalline local-field effects.}
\label{fig:imeps}
\end{figure}

In Fig.\ \ref{fig:imeps} the corresponding values of the so-called  energy-loss function
${\rm Im}\left[-\epsilon_{{\bf G},{\bf G}}^{-1}({\bf q},\omega)\right]$
are displayed. Our results, obtained for a small but finite 
momentum transfer, are in excellent agreement with the measured
dielectric function, both showing a double peak structure in 
the energy-loss function.
The solid line represents the solution when we consider the crystalline 
local-field effects appearing through the dependence of the diagonal 
elements of the interacting  response matrix $\chi_{{\bf G},{\bf G}'}({\bf
q},\omega)$ on the off-diagonal elements of the polarizability  $\chi^0_{{\bf
G},{\bf G}'}({\bf q},\omega)$.
The dashed-dotted line describes the behavior when this effect is not taken
into account. 
Even if the shape of both curves is quite similar, the solid line accounts
better for the experimental result than the dashed-dotted line, the 
difference being bigger for higher energies.
The small finite q-value used keeps all the features of the optical spectrum
and therefore can be directly compared with the measured one.

In order to investigate the role of localized $d$ bands in the dynamical 
response of gold, Fig.\ \ref{fig:imeps} also shows (dotted line) the 
energy-loss function calculated from an {\it ab initio} pseudopotential with 
the 5$d$ shell assigned to the core. The plot shows that a combined 
description of both localized 5$d$ bands and delocalized 6$s^{1}$ electrons 
is needed to correctly address the actual electronic response of gold.

For Au, as well as for the other noble metals, one expects to observe a 
resonance near 9 eV corresponding to the density of free electrons. The role 
of the Au $d$ bands is to provide a polarizable background which lowers
this free electron plasma frequency. We note from Fig.\ \ref{fig:epsilon}
that near 5 eV the real part of the dielectric function 
(${\rm Re}\,\epsilon$)
 is zero; however, the imaginary part (${\rm Im}\,\epsilon$)
is not small, due to the existence of interband transitions at these
energies which damp the free-electron plasmon. 

$d$-bands are also responsible, at higher energies, for a double peak 
structure in the energy-loss function, which arises from a combination
of band-structure effects and the building up of $d$-like collective modes. 
Since these peaks occur at energies ($\sim 25\,{\rm eV}$ and
$\sim 35\,{\rm eV}$) where ${\rm Re}\,\epsilon$ is nearly zero 
(see Fig.\ \ref{fig:epsilon}) 
they are in the nature of collective excitations, the small but finite value
of ${\rm Im}\,\epsilon$ at these energies accounting for the width of the 
peaks. A similar double-peak structure has been obtained for the other
noble metals and seems to be a general feature stemming from the particular
$d$-electron band structure (localization and hybridization with the 
delocalized $s$-electrons) of the three noble metals.

\begin{figure}
\epsfclipon
\epsfxsize=8cm 
\centerline{\psfig{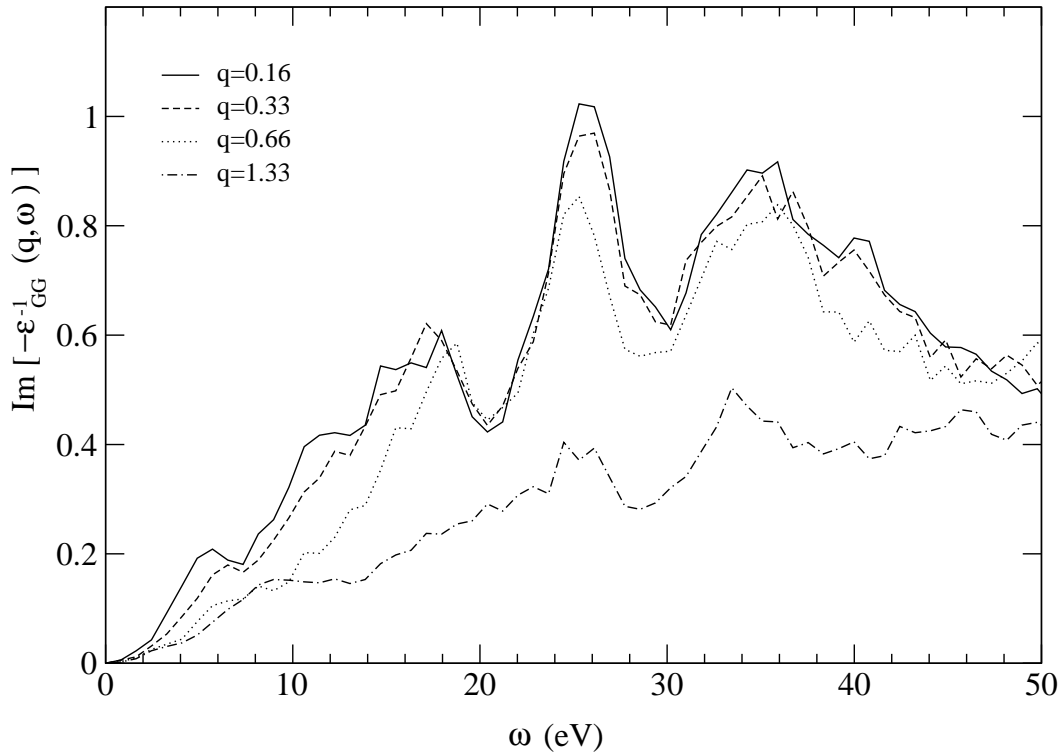}}
\caption{\protect\small  
RPA energy-loss function of Au along the (100) direction,
for various values of the momentum transfer
$|{\bf q}+{\bf G}|$: 0.16, 0.33, 0.66 and 1.33, in units of
$2\pi /a$ ($a=3.01 \AA$).}
\label{fig:q-epsilon}
\end{figure}

In order to see the dependence of the energy-loss function on the momentum
transfer ${\bf q}+{\bf G}$, we show in Fig.\ \ref{fig:q-epsilon}
the RPA dynamic structure factor for ${\bf G}=0$, as obtained for various 
values of $q$ along the (100) direction. 
As long as the
$5d$ shell of Au is assigned to the core, one finds a well-defined 
free-electron 
plasmon for wave vectors
up to the critical momentum transfer where the plasmon excitation 
enters the continuum of intraband particle-hole excitations. This
free-electron plasmon, which is known to show a characteristic 
positive dispersion with
wave vector, is found to be completely damped when a
realistic description of $5d$ orbitals is included in the calculations. At
higher energies and small momenta (q $<$ 0.7), 
$d$-like collective excitations originate a
double-peak structure which presents no dispersion. For larger values of the
momentum transfer ${\bf q}+{\bf G}$, single-particle excitations take over
the collective ones up to the point that above a given cutoff, the spectra
is completely dominated by the kinetic-energy term as expected. 

\section{Conclusions}
We have presented {\it ab initio} pseudopotential calculations of
the dynamical density-response function of Au, by including $d$-electrons as
part of the valence complex. In the long-wavelength limit (${\bf q}\to 0$),
$d$-bands provide a polarizable background that lowers the free-electron plasma
frequency. $d$-electrons are also responsible for a full damping of this
$s$-like collective excitation and for the appearance of a $d$-like double-peak
structure in the energy-loss function, in agreement with the experimentally
determined optical response of gold. We have analyzed the dependence of the
dynamical structure factor on the momentum transfer, and we have found that, 
for values of the momentum transfer over a given cutoff, a 
less-pronounced double-hump is originated by the existence of interband
electron-hole excitations. The dynamical response function in Cu and Au
shows very similar features, although in Au the peaks appear for higher
energies.\\

We acknowledge partial support by the Basque Hezkuntza, Unibertsitate 
eta Ikerketa Saila.

\end{document}